\documentclass[final,5p,times]{elsarticle}

\usepackage{url}
\usepackage{xcolor}
\usepackage{amssymb}
\usepackage{amsmath, mathtools}

\newcounter{bla}

\journal{Computer Physics Communications}

\begin{document}

\begin{frontmatter}

\title{SALMON 2.3: Implementation of divide-and-conquer ground-state initialization for large-scale real-time TDDFT}

\author[a]{Shunsuke Yamada\corref{author}}
\author[a]{Tomohito Otobe}

\cortext[author] {Corresponding author.\\\textit{E-mail address:} yamada.shunsuke@qst.go.jp}
\address[a]{Kansai Institute for Photon Science, National Institutes for Quantum Science
and Technology (QST), Kyoto 619-0215, Japan}

\begin{abstract}
In large-scale real-time time-dependent density functional theory (TDDFT) simulations, preparing the ground-state electronic structure can demand more computation than the subsequent time propagation. This creates a major bottleneck for simulations of non-equilibrium electron dynamics. This limitation is particularly severe for realistic systems, such as disordered materials, liquids, nanostructures, and heterogeneous condensed-matter systems, that contain thousands to tens of thousands of atoms. Real-time TDDFT provides a powerful framework for describing nonlinear and strong-field phenomena, including high-harmonic generation and light-induced phase transitions. However, its application to large-scale systems is hindered by the computational cost of conventional ground-state density functional theory (DFT) calculations.


SALMON is an open-source first-principles code for light-matter interaction simulations based on real-time TDDFT on real-space grids. It supports massively parallel calculations that combined message passing interface (MPI) with OpenMP or GPU acceleration [1,2]. 
In SALMON 2.3, a new stable version of the code, we implement a divide-and-conquer density functional theory (DC-DFT) scheme. We combine this scheme with a postprocessing method that reconstructs spatially extended Kohn--Sham orbitals of the entire system. 
These reconstructed global orbitals serve directly as initial states for the real-time TDDFT module of SALMON. 
This establishes a practical workflow that connects efficient ground-state preparation based on DC-DFT to the standard real-time, real-space TDDFT framework. Therefore, nonequilibrium and nonlinear phenomena in large-scale systems can be simulated at a realistic computational cost.

The present approach combines the efficiency of DC-DFT for ground-state preparation with the robustness and general applicability of conventional real-time TDDFT in SALMON. In particular, the self-consistent-field procedure based on DC-DFT exhibits linear scaling with system size. This directly addresses a major bottleneck in large-scale electron-dynamics simulations. We describe the computational procedure, parallelization strategy, and input/output design of the implementation. Weak-scaling measurements using Si supercells on Fugaku confirm the linear-scaling behavior of the DC-DFT implementation. We assess the accuracy of DC-initialized real-time TDDFT for a 512-atom amorphous Si system and a bulk H$_2$O liquid system containing 4134 atoms. 
These results demonstrate that the present workflow provides a practical route toward large-scale simulations of nonequilibrium electron dynamics by combining linear-scaling ground-state preparation with the established real-space, real-time TDDFT framework of SALMON.
\\
\noindent \textbf{NEW VERSION PROGRAM SUMMARY}

\begin{small}
\noindent
{\em Program Title: } SALMON: Scalable Ab-initio Light--Matter simulator for Optics and Nanoscience \\
{\em CPC Library link to program files:} (to be added by Technical Editor) \\
{\em Developer's repository link:} https://github.com/SALMON-TDDFT/SALMON2 \\
{\em Licensing provisions(please choose one):} Apache-2.0  \\
{\em Programming language:} Fortran 2003                        \\
{\em Journal reference of previous version:} Comput. Phys. Commun. \textbf{235}, 356--365 (2019).                \\
{\em Does the new version supersede the previous version?:} Yes   \\
{\em Reasons for the new version:} 
The new version adds a DC-DFT workflow to SALMON for efficient ground-state preparation of large systems. It also reconstructs global Kohn--Sham orbitals from fragment orbitals, so that DC-DFT results can serve directly as initial states for conventional real-time TDDFT simulations. \\
{\em Summary of revisions:} 
The main revisions include a DC-DFT calculation mode, new input keywords for fragment decomposition and buffer regions, and a DC-LCFO (linear combinations of fragment orbitals) postprocessing solver. The output structure has also been extended to store fragment data, basis functions, Hamiltonian matrices, and wavefunction coefficients. \\
{\em Nature of problem:} 
Real-time TDDFT requires accurate ground-state Kohn--Sham orbitals as initial states. For large systems, conventional ground-state DFT becomes expensive because diagonalization and orthonormalization scale poorly with system size. This cost can exceed that of subsequent time propagation, limiting large-scale electron-dynamics simulations. \\
{\em Solution method:} 
The system is divided into overlapping fragments, and local Kohn--Sham equations are solved independently. The total electron density is assembled from fragment contributions, yielding linear-scaling SCF calculations. Global Kohn--Sham orbitals are then reconstructed from fragment orbitals by DC-LCFO and used as initial states for real-time TDDFT. \\
{\em Additional comments, including restrictions and unusual features:}
The current implementation targets large periodic systems on real-space grids and assumes regular fragment decomposition. \(k\)-point sampling is not currently included in the DC-DFT mode. The final DC-LCFO diagonalization does not scale linearly; however, the reduced Hamiltonian dimension is much smaller than in conventional real-space calculations.
\\

\end{small}

\end{abstract}
\end{frontmatter}

\section{Introduction}

First-principles simulations based on density functional theory (DFT)~\cite{Hohenberg1964,Kohn1965} and time-dependent density functional theory (TDDFT)~\cite{Runge1984} are essential tools for studying electronic structures~\cite{Burke2012,Jones2015} and ultrafast electron dynamics~\cite{UllrichTDDFT,MarquesTDDFT,Sato2025} in condensed-matter systems. Real-time TDDFT is particularly powerful because it directly follows the nonequilibrium motion of electrons under external electromagnetic fields. This approach can describe nonlinear and strong-field phenomena beyond the scope of linear-response calculations. 

In practice, however, real-time TDDFT has been applied mainly to small and medium-sized systems. 
Simulations of very large systems, by contrast, have often been limited to ground-state or linear-response approaches~\cite{VandeVondele2012,Nakata2020,Prentice2020,Zuehlsdorff2013,ORourke2015,Bussy2021}. This leaves an important gap, as many realistic systems, such as disordered materials, amorphous structures, nanoscale heterogeneous systems, liquids, and interfaces, require large system sizes and an explicit real-time description of electron dynamics. Localized-basis approaches can provide efficient descriptions of electron dynamics induced by weak pulses close to the linear response regime ~\cite{ORourke2015,Prentice2020,Takimoto2007,Feng2025}. 
This picture breaks down into strongly excited, plasma-like states, where the number of basis functions needed to represent highly delocalized electronic states can become prohibitively large. For highly nonlinear and nonequilibrium phenomena in large systems, such as high-harmonic generation and light-induced phase transitions, real-time simulations must therefore rely on basis-independent frameworks, such as real-space finite-difference methods.

SALMON has been developed as an open-source computational framework for light-matter interaction simulations based on real-time TDDFT on real-space grids~\cite{Noda2019,SALMON_web}. By solving the time-dependent Kohn--Sham (KS) equation with a real-time, real-space finite-difference approach, SALMON can describe the coupled dynamics of electrons, electromagnetic fields, and atomic motion. It is designed for massively parallel execution using message passing interface (MPI) together with OpenMP or GPU acceleration~\cite{Noda2019,SALMON_web}. Large-scale real-time simulations have already been demonstrated with SALMON, for example, for amorphous SiO$_2$ thin films containing 10,224 atoms on Fugaku~\cite{Hirokawa2022}. Such calculations demonstrate the potential of large-scale real-time TDDFT. At the same time, they expose a central bottleneck: preparing the initial ground state with conventional DFT. In conventional real-space implementations, the computational cost of ground-state DFT scales as \(O(N^3)\) with the system size \(N\), mainly owing to diagonalization and orthonormalization of KS orbitals. Real-time TDDFT propagation, by comparison, scales only as \(O(N^2)\). Therefore, the ground-state calculation can become more expensive than the subsequent time propagation, posing a critical obstacle to extending real-time TDDFT to larger and more complex systems.

Linear-scaling DFT methods provide a natural route to overcome this bottleneck~\cite{Goedecker1999,Bowler2011}. Among them, the divide-and-conquer (DC) method~\cite{Yang1991,Yang1995,Shimojo2005,Shimojo2008,Ohba2012,Shimojo2014} is attractive because it exploits the locality, or nearsightedness, of electronic matter~\cite{Kohn1996,Prodan2006}. In DC-DFT, the system is divided into overlapping subsystems (fragments). Local KS equations are solved independently for each fragment, and the total electron density is assembled from their contributions. This reduces the cost of the self-consistent field (SCF) procedure to linear scaling with system size. However, this locality-based formulation is not directly compatible with conventional real-time TDDFT. Real-time propagation requires global KS orbitals extending over the whole simulation cell, whereas DC-DFT naturally provides only fragment-localized orbitals.

To bridge this gap, Ref.~\cite{Yamada2017} proposed a postprocessing method based on DC-DFT in which spatially extended global KS orbitals are reconstructed from fragment orbitals. In this method, a reduced orthonormal basis is constructed from low-energy fragment orbitals and the Hamiltonian matrix represented in this basis is diagonalized to obtain global electronic states. The resulting whole-system orbitals are expressed as linear combinations of fragment orbitals (LCFO), enabling a practical reconstruction of extended KS orbitals for large systems. We refer to this reconstruction scheme as DC-LCFO. The method was demonstrated for systems such as P-doped Si and InGaN/GaN superlattices, reproducing eigen energies and extended wavefunctions with useful accuracy~\cite{Yamada2017}. 
Nevertheless, its integration into a real-time TDDFT code as a practical initial-state preparation workflow had not been established.

In this paper, we implement DC-DFT and DC-LCFO solvers in SALMON and integrate them with the real-space, real-time TDDFT workflow. 
This implementation makes it possible to simulate nonlinear and nonequilibrium phenomena in large-scale systems at a realistic computational cost. 
The DC-DFT mode is added to the ground-state module of SALMON and operates much like conventional DFT, requiring only a few additional input parameters. 
After the DC-DFT SCF calculation, the DC-LCFO solver reconstructs global KS orbitals from fragment orbitals and passes them to the standard real-time, real-space TDDFT module of SALMON. 
Although DC-LCFO contains an $O(N^3)$ diagonalization step, its cost is modest in the overall workflow because it is executed only once as a postprocessing procedure.

The present workflow establishes a practical approach to large-scale simulations of nonequilibrium electron dynamics by combining linear-scaling ground-state preparation with the established real-space, real-time TDDFT framework of SALMON. 
The key advantage of this strategy is that it considerably reduces the cost of initial-state preparation for large systems. At the same time, it retains the robustness and broad applicability of the real-space, real-time TDDFT propagation scheme, including strong-field simulations. 
In particular, the present workflow enables microscopic simulations of coupled electron, electromagnetic-field, and atomic dynamics in large systems, as demonstrated in Ref.~\cite{Hirokawa2022}. This capability allows simulations that more closely reflect realistic conditions in optical-science experiments.
The reconstructed orbitals can also serve as initial orbitals for subsequent conventional DFT calculations, substantially reducing the number of SCF iterations required when higher ground-state accuracy is needed.

To demonstrate the capability of the implementation, we examine performance and numerical accuracy. Weak-scaling tests on Fugaku show the linear-scaling behavior of the DC-DFT SCF procedure. We then assess the accuracy of real-time TDDFT calculations initialized with DC-DFT for a 512-atom amorphous Si system. We compare the real-time current, excitation energy, linear-response dielectric function, and high-harmonic spectra against those obtained from conventional DFT initial states. As a more practical demonstration of large-system real-time TDDFT, we perform high-harmonic generation calculations for a bulk H$_2$O liquid system containing 4,134 atoms. These tests show that the present workflow provides practically useful accuracy for large-scale electron-dynamics simulations. They further clarify that the remaining discrepancies mainly originate from the quality of the DC-DFT ground-state density rather than from the LCFO reconstruction itself.

The remainder of this paper is organized as follows. 
Section~\ref{sec:methods} describes the DC formalism adopted in SALMON and summarizes the DC-LCFO method. 
Section~\ref{sec:results} presents performance benchmarks and accuracy tests for representative systems, including amorphous Si and liquid water systems under linear and nonlinear optical excitation. 
Finally, Section~\ref{sec:conclusion} summarizes the main findings and discusses future extensions of the present framework.

\section{Methods}
\label{sec:methods}

\subsection{Overview of the divide-and-conquer workflow in SALMON}

To accelerate ground-state preparation for large-scale real-time TDDFT simulations, we implement a DC-DFT workflow in SALMON. 
The overall workflow consists of the following steps: (i) an SCF loop based on DC-DFT for the total ground-state electron density, (ii) reconstruction and storage of the global KS orbitals by DC-LCFO, and (iii) conventional real-time TDDFT (or ground-state DFT) calculations using the reconstructed orbitals as initial states. 
The DC-DFT part follows the standard formulation and relies on the nearsightedness principle of the electron density in DFT~\cite{Kohn1996,Prodan2006}. 
By exploiting the short-range character of the electron density, the system is divided into small subsystems or fragments. The total electron density is then reconstructed from these fragments, achieving linear-scaling computational cost. 
DC-DFT is especially effective for gapped and disordered systems, in which the correlations in the density matrix decay exponentially~\cite{Prodan2006}.

A key requirement for coupling DC-DFT to conventional real-time TDDFT is the availability of global KS orbitals as initial states. 
For this purpose, we adopt the fragment-orbital-based postprocessing method proposed in Ref.~\cite{Yamada2017}, which we refer to as DC-LCFO. 
In this method, a reduced orthonormal basis is constructed from low-energy fragment orbitals, a compact Hamiltonian matrix for the whole system is assembled in this basis, and the global KS orbitals are reconstructed from the resulting eigenvectors. 
In SALMON, these reconstructed KS orbitals are directly passed to the standard real-time TDDFT propagation module without modifying the time-evolution formalism itself. 
Furthermore, the reconstructed orbitals can serve as initial orbitals for conventional DFT calculations, accelerating SCF convergence. 
The details of these workflows are described in the following subsections.

\subsection{Divide-and-conquer ground-state calculation}

In the DC-DFT formalism \cite{Yang1991,Yang1995,Shimojo2014,Yamada2017}, the physical space of the total system is represented as a union of nonoverlapping core domains $\{\Omega^\alpha_0\}$. Each core domain is then extended by a buffer region to define an overlapping fragment $\Omega^\alpha$. 
For each fragment $\alpha$, the fragment KS equation is solved as follows:
\begin{equation}
\hat{H}^{\alpha}\phi_i^{\alpha}(\mathbf{r}) = \varepsilon_i^{\alpha}\phi_i^{\alpha}(\mathbf{r}), \label{eq:KS}
\end{equation}
where $\hat{H}^{\alpha}$ denotes the local KS Hamiltonian defined in $\Omega^\alpha$.
$\phi_i^{\alpha}$ and $\varepsilon_i^{\alpha}$ denote the corresponding fragment orbitals and eigenvalues, respectively. 

\begin{figure}
    \centering
    \includegraphics[keepaspectratio,width=8cm]{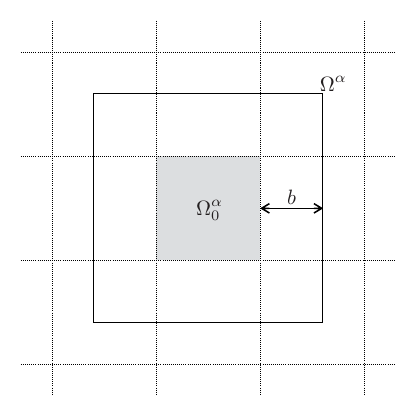}
    \caption{\label{fig:frag}
    Schematic of a fragment $\Omega^{\alpha}$ in the divide-and-conquer method. 
    The shaded area denotes the core region $\Omega^\alpha_0$, and $b$ is the buffer thickness.
    }
\end{figure}

In the present SALMON implementation, the whole system is assumed to be a rectangular cell with periodic boundary conditions. 
Because this implementation targets large periodic condensed-matter systems, $k$-point sampling is not
currently included.
The core regions are defined by dividing the whole system into a regular grid according to the prescribed number of fragments. Each fragment is then constructed by attaching a buffer region of thickness $b$ to the corresponding core region (Fig.~\ref{fig:frag}). 
Periodic boundary conditions are imposed at the boundaries of each fragment, $\partial \Omega^\alpha$.
The total electron density, $\rho_{\rm tot}(\mathbf{r})$, is reconstructed by summing the fragment densities projected onto the respective core regions, $\bar{\rho}_{\alpha}(\mathbf{r})$.
These quantities are defined as follows:
\begin{equation}
    \rho_{\rm tot}(\mathbf{r})=\sum_{\alpha}\bar{\rho}_{\alpha}(\mathbf{r}),
\end{equation}
\begin{equation}
\bar{\rho}_{\alpha}(\mathbf{r})=
    \begin{cases}
    \rho_{\alpha}(\mathbf{r}), & (\mathbf{r} \in \Omega^\alpha_0) \\
    0,                 & (\mathbf{r} \notin \Omega^\alpha_0) ,  
    \label{eq:rho_proj}
  \end{cases}
\end{equation}
\begin{equation}
    \rho_{\alpha}(\mathbf{r})=\sum_i f(\varepsilon_i^{\alpha}-\mu)|\phi_i^{\alpha}(\mathbf{r})|^2 ,
\end{equation}
where $\rho_{\alpha}(\mathbf{r})$ is the fragment density and $f$ is the Fermi-Dirac distribution function.
The chemical potential $\mu$ is determined globally to conserve the total number of electrons, $N_e=\int d^3r\,\rho_{\rm tot}(\mathbf{r})$. 
The fragment KS Hamiltonian $\hat{H}^{\alpha}$ is defined as follows:
\begin{equation}
    \hat{H}^{\alpha}=-\frac{\hbar^2}{2m}\nabla^2 + V_{\rm loc}(\hat{\mathbf{r}})+\hat{V}_{\rm NL}+V_{\rm H}[\rho_{\rm tot}](\hat{\mathbf{r}}) + V_{\rm xc}[\rho_{\alpha}](\hat{\mathbf{r}})+V^{\alpha}(\hat{\mathbf{r}}),
\end{equation}
where $V_{\rm loc}$ and $\hat{V}_{\rm NL}$ denote the local and nonlocal parts, respectively, of the ionic pseudopotential defined within $\Omega^{\alpha}$.
$V_{\rm H}[\rho_{\rm tot}]$ denotes the Hartree potential, determined from the global Poisson equation using the total density $\rho_{\rm tot}(\mathbf{r})$.
$V_{\rm xc}[\rho_{\alpha}]$ denotes the exchange-correlation potential calculated from the fragment density $\rho_{\alpha}(\mathbf{r})$.
$V^{\alpha}(\mathbf{r}) = [\rho_{\alpha}(\mathbf{r}) - \rho_{\rm tot}(\mathbf{r})]/\xi$ with $\xi > 0$ denotes the density-template potential \cite{Ohba2012,Shimojo2014} and is set to zero by default.
The SCF loop for the total density $\rho_{\rm tot}(\mathbf{r})$ is iterated using the same procedure as in conventional DFT but solving the localized KS equations Eq.~(\ref{eq:KS}) for the respective fragments. 

The dominant computational cost in a ground-state calculation arises from the diagonalization of the KS equation and the orthonormalization of the KS orbitals. 
In DC-DFT, however, only fragment-localized KS orbitals are considered. 
Therefore, when the system size is increased, the fragment size can be kept constant. Only the number of fragments needs to increase, so that the computational cost scales only with the number of fragments.
Although the calculation of the Hartree potential requires a fast Fourier transform (FFT) over the whole system, its computational cost scales as $O(N{\rm log}N)$. 
Therefore, the overall cost of the DC-DFT calculation is approximately $O(N)$, namely, it exhibits linear scaling~\cite{Goedecker1999,Bowler2011,Shimojo2014}.

For practical use in SALMON, fragment decomposition is controlled by user-defined input parameters specifying the number of fragments in each direction, the number of grid points in the buffer region, and the number of fragment states to be retained. 
In addition, density summation can be performed at a finite electronic temperature for the occupation function $f$. 
In this paper, a temperature of 300~K is used for the density sum in the representative calculations for numerical stability.
The norm-conserving pseudopotential~\cite{Troullier1991} and (adiabatic) local density approximation~\cite{Perdew1981} are used in the calculations.

\subsection{Reconstruction of global Kohn--Sham orbitals}

After the SCF loop has converged, the global KS states are reconstructed using the DC-LCFO postprocessing method, which employs a compact basis derived from fragment orbitals.
Following the procedure of Ref.~\cite{Yamada2017}, we first select fragment orbitals whose eigenvalues are lower than the prescribed cutoff energy $\varepsilon_{\mathrm{cut}}$. 
To suppress redundant contributions from the buffer region, each selected fragment orbital is projected onto the corresponding core region $\Omega^{\alpha}_0$, yielding projected orbitals $\{\bar{\phi}_i^\alpha\}$ defined in the same manner as Eq.~(\ref{eq:rho_proj}). 
An overlap matrix is then constructed independently for each fragment.
\begin{equation}
S_{ij}^{\alpha} = \langle \bar{\phi}_i^\alpha | \bar{\phi}_j^\alpha \rangle .
\end{equation}
The overlap matrix is diagonalized. Eigenvectors associated with sufficiently small eigenvalues are then discarded using a threshold parameter $\lambda_{\mathrm{cut}}$, which removes overcompleteness and controls the number of basis functions \cite{Yamada2017}.

The retained eigenvectors of the overlap matrix define a new set of orthonormal basis functions $\{\lambda_i^\alpha\}$, each localized to a given core domain $\Omega_0^\alpha$ but built from low-energy fragment orbitals. 
The basis functions are localized in the core region of each fragment and orthonormalized within it.
Therefore, they also satisfy the orthonormality condition as a whole:
\begin{equation}
    \langle \lambda_i^\alpha| \lambda_{i'}^{\alpha'}\rangle = \delta_{\alpha\alpha'}\delta_{ii'}.
\end{equation}
In this way, the basis is compact and systematically generated from fragment electronic structure data, while still capable of representing spatially extended states of the whole system through linear combinations across fragments.
The use of a reduced basis is essential for avoiding the large Hamiltonian dimension associated with conventional real-space or plane-wave representations. 
In the original DC-LCFO formulation, practical accuracy was achieved with roughly 10--20 basis functions per atom in representative systems. The present SALMON implementation follows the same construction principle \cite{Yamada2017}.
In the validation examples presented here, representative values such as $\lambda_{\mathrm{cut}} = 10^{-7}$ and $\varepsilon_{\mathrm{cut}}-\mu = 0.1$~eV are used; the optimal choice, however, may depend on the target system and the desired energy window.

To obtain the global KS orbitals, we construct the Hamiltonian matrix $\langle \lambda_i^\alpha|\hat{H}| \lambda_{i'}^{\alpha'}\rangle$ of the whole system using the basis functions $\{\lambda_i^\alpha\}$ and then diagonalize the matrix.
Here, $\hat{H}$ is the KS Hamiltonian defined in the whole system:
\begin{equation}
    \hat{H}=-\frac{\hbar^2}{2m}\nabla^2 + V_{\rm loc}(\hat{\mathbf{r}})+\hat{V}_{\rm NL}+V_{\rm H}[\rho_{\rm tot}](\hat{\mathbf{r}}) + V_{\rm xc}[\rho_{\rm tot}](\hat{\mathbf{r}}).
\end{equation}
In practice, the total KS Hamiltonian $\hat{H}$ is approximated by the local KS Hamiltonian $\hat{H}^{\alpha}$, and the matrix elements are evaluated by applying the ordinary Hamiltonian operation to the basis functions, $\lambda_i^\alpha({\mathbf{r}})\rightarrow \hat{H}^{\alpha} \lambda_i^\alpha({\mathbf{r}})$, as follows:
\begin{equation}
    \langle \lambda_i^\alpha|\hat{H}| \lambda_{i'}^{\alpha'}\rangle=
    \begin{cases}
        \langle \lambda_i^\alpha|\hat{H}^{\alpha}| \lambda_{i'}^{\alpha}\rangle, & ({\alpha}={\alpha'}), \\
        \frac{1}{2} \left[ (\langle \lambda_i^\alpha|\hat{H}^{\alpha})| \lambda_{i'}^{\alpha'}\rangle + 
        \langle \lambda_i^\alpha|(\hat{H}^{\alpha'}| \lambda_{i'}^{\alpha'}\rangle) \right] , & ({\alpha}\neq {\alpha'}).
    \end{cases}
\end{equation}
The resulting Hamiltonian matrix is sparse in the sense that nonzero couplings arise between basis functions associated with the same fragment or with neighboring fragments, reflecting the semilocality of the KS Hamiltonian.
Here, the second line of the right-hand side involves inner products between vectors belonging to different fragments. In the parallel implementation, this requires one-to-one communication between the MPI processes responsible for the buffer region and the overlapping neighboring core region.

Once the Hamiltonian matrix has been diagonalized, we obtain eigenvectors expressed in the orthonormal basis $\{\lambda_i^\alpha\}$, from which the global KS orbitals are reconstructed as follows:
\begin{equation}
\psi_n(\mathbf{r}) = \sum_{\alpha}\sum_{i \in \alpha} C_{\alpha i,n}\,\lambda_i^\alpha(\mathbf{r}),
\end{equation}
where $C_{\alpha i,n}$ are the coefficients of the $n$-th eigenstate on the reduced basis.
This reconstruction yields orbitals that are defined on the global real-space grid and suitable for direct use in subsequent conventional DFT or TDDFT calculations within SALMON.

The basis functions $\lambda_i^\alpha({\mathbf{r}})$ and wavefunction coefficients $C_{\alpha i,n}$ are written to output files that the subsequent simulation stages can read. 
In the current directory structure, the DC-related data are separated into directories for each fragment, which contain basis-function data, local Hamiltonian data, and wavefunction coefficients. 

The matrix construction includes communications among neighboring fragments when off-diagonal blocks are required, while dominant local operations are carried out independently on each fragment. 
According to the implemented workflow, the SCF loop for the total density, the basis construction, and the Hamiltonian assembly are all treated as nearly linear-scaling computations.
However, the final matrix diagonalization remains an $O(N^3)$ computational step, although the use of a compact basis substantially decreases this dimension compared to conventional calculations. 
This division of computational cost is a central feature of the present implementation.
Moreover, since diagonalization is performed only once as a postprocessing step, its contribution to the overall computational cost is not a major concern in practice.
The current implementation uses EigenExa~\cite{Fukaya2015,EigenExa} for diagonalization, and a sparse-matrix solver, such as the Sakurai-Sugiura method~\cite{Sakurai2003}, is considered a future extension.

\subsection{Real-time TDDFT}

Once the global KS orbitals have been constructed using DC-DFT and its postprocessing method, DC-LCFO, we perform real-time TDDFT calculations using them as initial states. 
Details of the real-time TDDFT calculation are given in Ref.~\cite{Noda2019}. 
Taking $\psi_n(\mathbf{r})$ as the initial state of the time-dependent KS orbital $\psi_n(\mathbf{r},t)$, we solve the following time-dependent KS equation:
\begin{equation}
    i\hbar\frac{\partial}{\partial t}\psi_n(\mathbf{r},t)=\hat{H}(t)\psi_n(\mathbf{r},t),
\end{equation}
\begin{eqnarray}
    \hat{H}(t)&=&\frac{1}{2m}\left(-i\hbar\nabla +\frac{e}{c}\mathbf{A}(t) \right)^2 + V_{\rm loc}(\hat{\mathbf{r}}) \nonumber \\
    && + e^{-i\frac{e}{\hbar c}\mathbf{A}(t)\cdot \hat{\mathbf{r}}}\hat{V}_{\rm NL} e^{i\frac{e}{\hbar c}\mathbf{A}(t)\cdot \hat{\mathbf{r}}}  
    +V_{\rm H}[\rho_{\rm tot}(t)](\hat{\mathbf{r}}) \nonumber \\
    && + V_{\rm xc}[\rho_{\rm tot}(t)](\hat{\mathbf{r}}),
\end{eqnarray}
\begin{equation}
    \rho_{\rm tot}(\mathbf{r},t)=\sum_n^{\rm occ}|\psi_n(\mathbf{r},t)|^2,
\end{equation}
where $\mathbf{A}(t)$ is the spatially uniform vector potential of the applied electric field.
In this paper, we use the waveform given by
\begin{eqnarray}
{\bf A}(t) = 
\begin{cases}
    - \frac{c E_0}{\omega} \, \sin \left[ \omega \left( t
  - \frac{T}{2} \right) \right] \,\, \sin^2 \left( \frac{\pi t}{T}
\right) \hat{\bf z} , & (0<t<T), \\
0, & (\rm{otherwise}),
\end{cases}
\label{eq:pulse}
\end{eqnarray}
where $E_0$, $\omega$, and $T$ are the peak amplitude, central frequency, and pulse duration of the laser pulse, respectively.

The main physical quantities analyzed in real-time TDDFT are the excitation energy and electron current density.
The former is given by 
\begin{equation}
    E_{\rm ex}(t)=E[\rho_{\rm tot}(t)]-E[\rho_{\rm tot}(t=0)],
\end{equation}
where $E[\rho]$ is the total energy functional of the electron density $\rho({\mathbf r})$.
The electron current density averaged over the calculation cell, $\mathbf{j}(t)$, is given by
\begin{equation}
 \mathbf{j}(t)=\int_{\Omega} \frac{d^3 r}{\Omega} \sum_n^{\rm occ} \psi^{\ast}_n(\mathbf{r},t) \frac{1}{i\hbar} \left[  \hat{\mathbf{r}}, \hat{H}(t) \right] \psi_n(\mathbf{r},t),
\end{equation}
where $\Omega$ is the volume of the cell.
These physical quantities are used in the latter part of this paper to assess the accuracy of the initial states prepared by DC-DFT.

\subsection{User input and output structure}

SALMON has dedicated input keywords for the DC-DFT workflow in the \path{&calculation} and \path{&dc} namelists. 
The main keyword is \path{yn_dc} = `\path{y/n}' (yes or no) in the \path{&calculation} namelist.
DC-DFT is activated when \path{yn_dc} = `\path{y}' is specified.
The \path{&dc} namelist contains keywords dedicated to DC-DFT, as listed in Table.~\ref{tab:dc_namelist}.

\begin{table*}[htbp]
\centering
\caption{Input parameters in the \texttt{\&dc} namelist for DC-DFT calculations in SALMON.}
\label{tab:dc_namelist}
\begin{tabular}{llp{10cm}}
\hline
Keyword & Type & Description  \\
\hline
\path{num_fragment(1:3)}
& Integer
& Number of fragment decompositions in each direction. \\
\path{nstate_frag}
& Integer
& Number of orbitals in each fragment. \\
\path{num_rgrid_buffer(1:3)}
& Integer
& Number of grid points corresponding to the buffer thickness $b$ in each direction. \\
\path{nproc_rgrid_tot(1:3)}
& Integer
& Number of MPI process decompositions in each direction for the real-space grid of the whole system. \\
\path{yn_dc_lcfo}
& Character(1)
& Keyword to activate the DC-LCFO calculation (`y' or `n'). \\
\path{lambda_cut}
& Real(8)
& Truncation threshold for overlap matrix eigenvalues $\lambda_{\rm cut}$. \\
\path{energy_cut}
& Real(8)
& Energy cutoff for fragment orbitals $\varepsilon_{\rm cut}$. \\
\hline
\end{tabular}
\end{table*}

The implementation is designed for massively parallel environments using the MPI library.
In the SCF loop of DC-DFT, the fragments are distributed over MPI processes, and additional parallelization can be introduced within each fragment through orbital and real-space decompositions. 
Parallel execution in each fragment is controlled through parameters of \path{nproc_k} ($k$-point MPI distribution; must be set to 1 for DC-DFT), \path{nproc_ob} (orbital MPI distribution), and \path{nproc_rgrid(1:3)} (real-space grid MPI distribution) in the \path{&parallel} namelist.
These input parameters are common to conventional calculations.

In the SALMON input structure, the MPI process distribution is related to fragment decomposition and internal parallelization.
\begin{eqnarray}
\#(\text{Processes})
&=&
\prod \texttt{nproc\_rgrid\_tot} \nonumber \\
&=&
\left(\prod \texttt{num\_fragment}\right) \nonumber \\
&\times &  \texttt{nproc\_ob} \times \left(\prod \texttt{nproc\_rgrid}\right),
\end{eqnarray}
where the left-hand side is the total number of MPI processes.
Here, the right-hand side of the first equality corresponds to the MPI process distribution for FFT of the global Poisson equation.
The right-hand side of the second equality is the MPI process distribution for the fragment KS orbitals, where $\prod$ \path{num_fragment} equals the number of fragments.

The output files are organized under a dedicated DC-DFT data directory (\path{data_dcdft}), which contains separate subdirectories for fragment data (\path{data_dcdft/fragments/000001}, \path{data_dcdft/fragments/000002}, $\cdots$) and total-system data (\path{data_dcdft/total}). 
The total-system directory stores quantities, such as the eigenenergies of the whole system calculated in the DC-LCFO solver.
Each fragment directory contains basis functions $\lambda_i^\alpha({\mathbf{r}})$ (\path{basis_functions.bin}), local Hamiltonian matrices $\langle \lambda_i^\alpha|\hat{H}| \lambda_{i'}^{\alpha'}\rangle$ (\path{hamiltonian_local.bin}), wavefunction coefficients $C_{\alpha i,n}$ (\path{wavefunctions.bin}), and grid-index information (\path{rgrid_index.bin}). 

To read files generated by DC-LCFO and perform conventional DFT or TDDFT calculations, the input keyword \path{yn_conventional_from_dcdft} = `\path{y}' should be specified in the \path{&calculation} namelist.
In this case, all other input keywords can be the same as those used for standard DFT or TDDFT calculations.

An advantage of representing the global KS orbitals in the DC-LCFO basis is that the data size can be significantly reduced. This is in contrast with wavefunction data represented on the global real-space grid used in conventional calculations within SALMON. 
In the latter case, the data size scales with both the number of real-space grid points and the number of orbitals, resulting in extremely large files for large systems. 
In the DC-LCFO representation, by contrast, the data size of the basis functions scales only with the number of fragments. The number of wavefunction coefficients, meanwhile, scales with the product of the number of fragments and the number of orbitals. 
Consequently, the overall data size in the DC-LCFO representation is much smaller than that in the real-space grid representation.

\subsection{Practical usage}

The DC-DFT workflow implemented in SALMON supports the following three practical use cases.

(1) The ground-state KS orbitals calculated in the DC-DFT mode (\path{yn_dc} = `\path{y}') are stored and then read as the initial states for conventional real-time TDDFT calculations (\path{yn_conventional_from_dcdft} = `\path{y}'). 
This approach is particularly useful for analyzing physical quantities in real time during excitation dynamics induced by intense laser irradiation in large-scale systems.
Because residual noise arises from the density error specific to DC-DFT, its applicability is limited for analyses in the linear-response regime and for high-harmonic spectra (see Sec.~\ref{sec:aSi}).

(2) The ground-state KS orbitals calculated in the DC-DFT mode (\path{yn_dc} = `\path{y}') are stored and then read as the initial states for a conventional ground-state DFT calculation (\path{yn_conventional_from_dcdft} = `\path{y}'). 
Because the initial wavefunctions are already nearly converged with respect to the SCF procedure, the subsequent convergence is extremely fast. 
In particular, for systems with a large gap, practically sufficient accuracy can be reached within only a few iterations. 
Therefore, compared with the standard workflow of performing DFT followed by TDDFT, a three-step workflow of DC-DFT, DFT, and TDDFT can substantially reduce the computational time of the DFT stage. Crucially, it does so while maintaining the same final accuracy.

(3) Only the electron density $\rho_{\rm tot}(\mathbf{r})$ obtained by a conventional DFT calculation is stored in a file. A DC-DFT calculation is then performed with the electron density fixed to the value read from this file (fixed-density DC).
In this case, the conventional DFT calculation does not need to output the extremely large wavefunction files. Instead, the global KS orbitals are reconstructed in the DC-LCFO basis, whose file size is much smaller. 
Because the electron density is fixed to that of conventional DFT, the density error specific to DC-DFT is eliminated. As a result, TDDFT calculations can be performed with essentially the same accuracy as when orbitals from conventional DFT are used as the initial states.
This approach is particularly useful when disk space usage must be minimized.

\subsection{Comparison with other approaches}

Representative real-time TDDFT codes besides SALMON include Octopus~\cite{Andrade2012}, Elk~\cite{Elk}, CP2K~\cite{Hanasaki2025}, and NWChem~\cite{Lopata2011}.
ONETEP~\cite{ORourke2015,Prentice2020} and SIESTA~\cite{Takimoto2007} are major linear-scaling DFT codes. Both have developed real-time TDDFT schemes based on localized orbitals.
To our knowledge, no code has yet been established that connects a linear-scaling DFT workflow to a standard real-time TDDFT scheme via the explicit reconstruction of global KS orbitals on a real-space grid.

A recent study~\cite{Feng2025} reported electron dynamics calculations for million-atom systems using real-time TDDFT within a discontinuous-Galerkin framework~\cite{Lin2012,Zhang2017}. 
Although that work and the present study both use compact localized basis functions, the underlying strategy differs. 
In the present implementation, DC-LCFO serves as a postprocessing step for the linear-scaling ground-state calculation based on DC-DFT, generating global KS orbitals on the real-space grid. 
These orbitals are then passed to the standard real-time, real-space finite-difference TDDFT scheme of SALMON, whereas the discontinuous-Galerkin method propagates the dynamics using a Hamiltonian matrix represented on a localized basis~\cite{Feng2025}. 
This difference is particularly important in the strong-field regime, where localized-basis TDDFT requires careful checks of basis convergence. Basis truncation can affect physical quantities there, especially for highly excited, delocalized, or plasma-like electronic states. 
Real-time, real-space finite-difference TDDFT has already proven successful for intense ultrashort pulses and high-harmonic generation \cite{Noda2019,Sato2025}.
Therefore, the present SALMON implementation offers a distinct route, combining an $O(N)$ initial-state preparation for very large systems with a grid-based TDDFT formalism suitable even for strong-field simulations.

\section{Results and Discussion}
\label{sec:results}

\subsection{Weak-scaling performance of the SCF loop in DC-DFT}

\begin{figure}
    \centering
    \includegraphics[keepaspectratio,width=8cm]{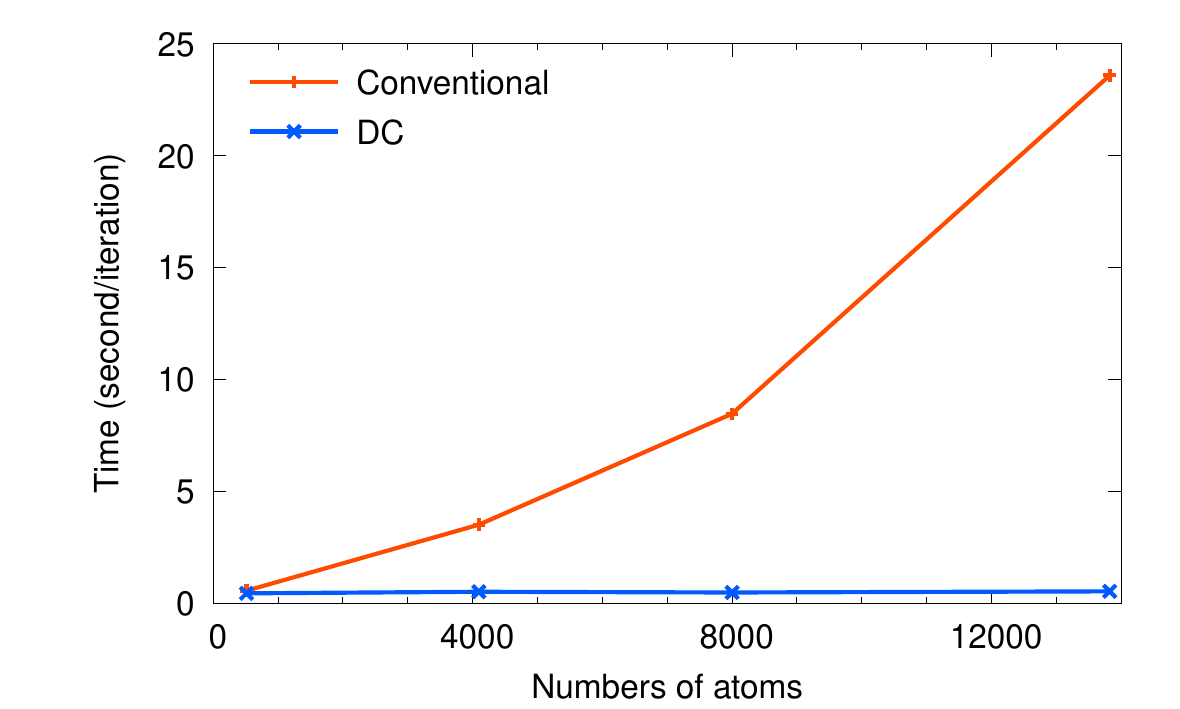}
    \caption{\label{fig:perf}  
    Comparison of computational time in weak-scaling calculations on Fugaku. 
    For Si supercells, one node is used per unit cell. 
    The red and blue lines denote conventional DFT and DC-DFT, respectively.
    }
\end{figure}

We begin by assessing the parallel performance of the DC-DFT SCF procedure implemented in SALMON.
As a representative benchmark, we consider Si supercells and distribute the calculation over the Fugaku supercomputer, using one node per unit cell. The cubic unit cell contains eight Si atoms in the diamond structure.
Figure~\ref{fig:perf} shows the calculation time per iteration step of SCF as a function of the number of atoms.
The red and blue lines correspond to results from conventional DFT and DC-DFT, respectively.

The calculations are performed using $n \times n \times n$ computational nodes of Fugaku for a $n \times n \times n$ supercell comprising $8n^3$ atoms.
The largest case is a $12 \times 12 \times 12$ supercell containing 13,824 atoms, using 1,728 Fugaku nodes.
We use 4 MPI processes and 12 OpenMP threads per node.
The grid spacing of the real-space grid is set to 0.45 {\AA}.
For the conventional DFT case, \path{nproc_ob} = 32 and \path{nproc_rgrid} = $(n/2,n/2,n/2)$ are specified.
For the DC-DFT case, the $n \times n \times n$ supercell is divided into $n \times n \times n$ fragments (\path{num_fragment} = $n,n,n$).
Here, \path{nproc_ob} = 1, \path{nproc_rgrid} = $(1,2,2)$, and \path{nproc_rgrid_tot} = $(n, 2n, 2n)$ are specified.
The buffer thickness $b$ is set to one Si lattice constant (5.43 {\AA}), and each fragment corresponds to a $3 \times 3 \times 3$ cell system (\path{num_rgrid_buffer} = $12,12,12$).

For the DC-DFT case, the measured wall-clock time per SCF iteration exhibits linear-scaling behavior over the tested system sizes.
By contrast, the conventional DFT result exhibits the expected $O(N^3)$ behavior.
The crossover between the two cases occurs at 512 atoms.
These results support the linear-scaling character of the DC-DFT SCF loop discussed in Sec.~\ref{sec:methods} and demonstrate that the implementation is suitable for large-scale execution on massively parallel architectures.

\subsection{Amorphous Si}
\label{sec:aSi}

\begin{figure}
    \centering
    \begin{tabular}{c}
    \includegraphics[keepaspectratio,width=8cm]{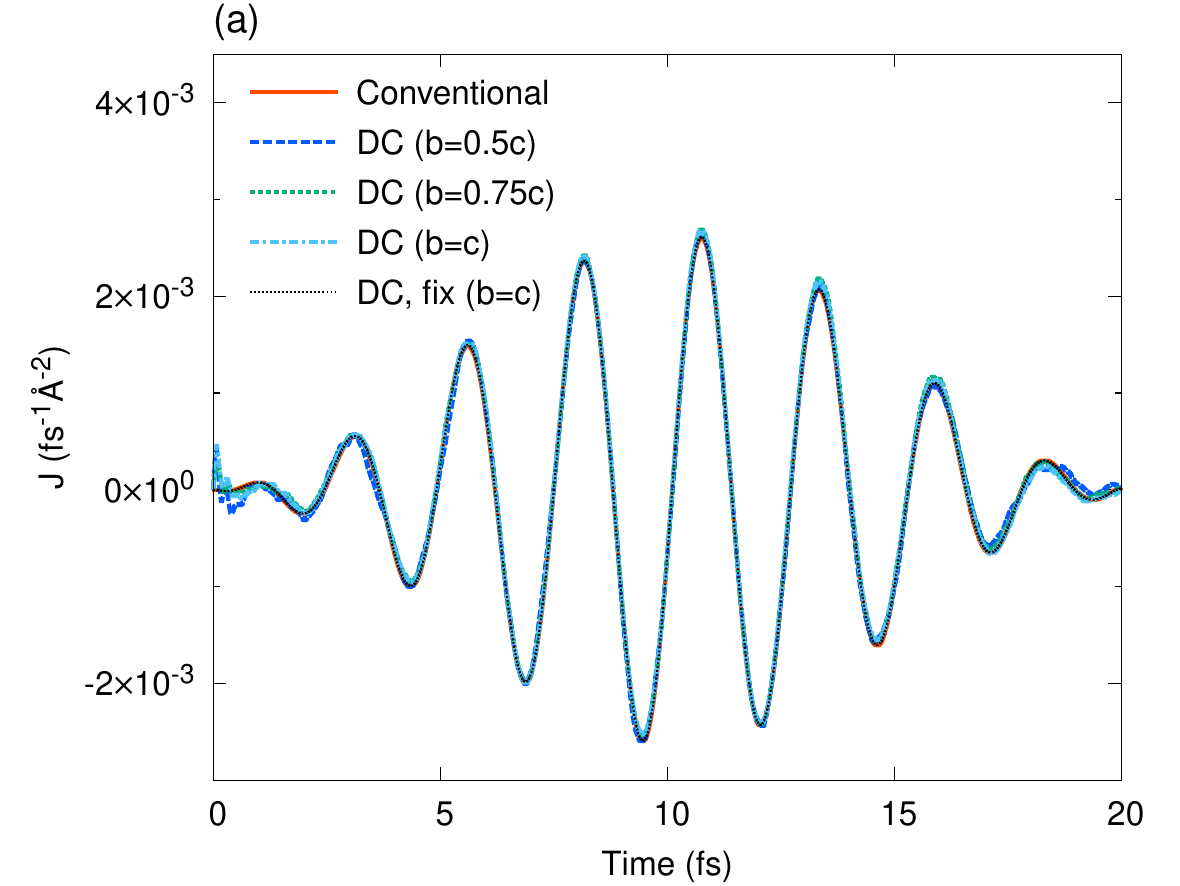}\\
    \includegraphics[keepaspectratio,width=8cm]{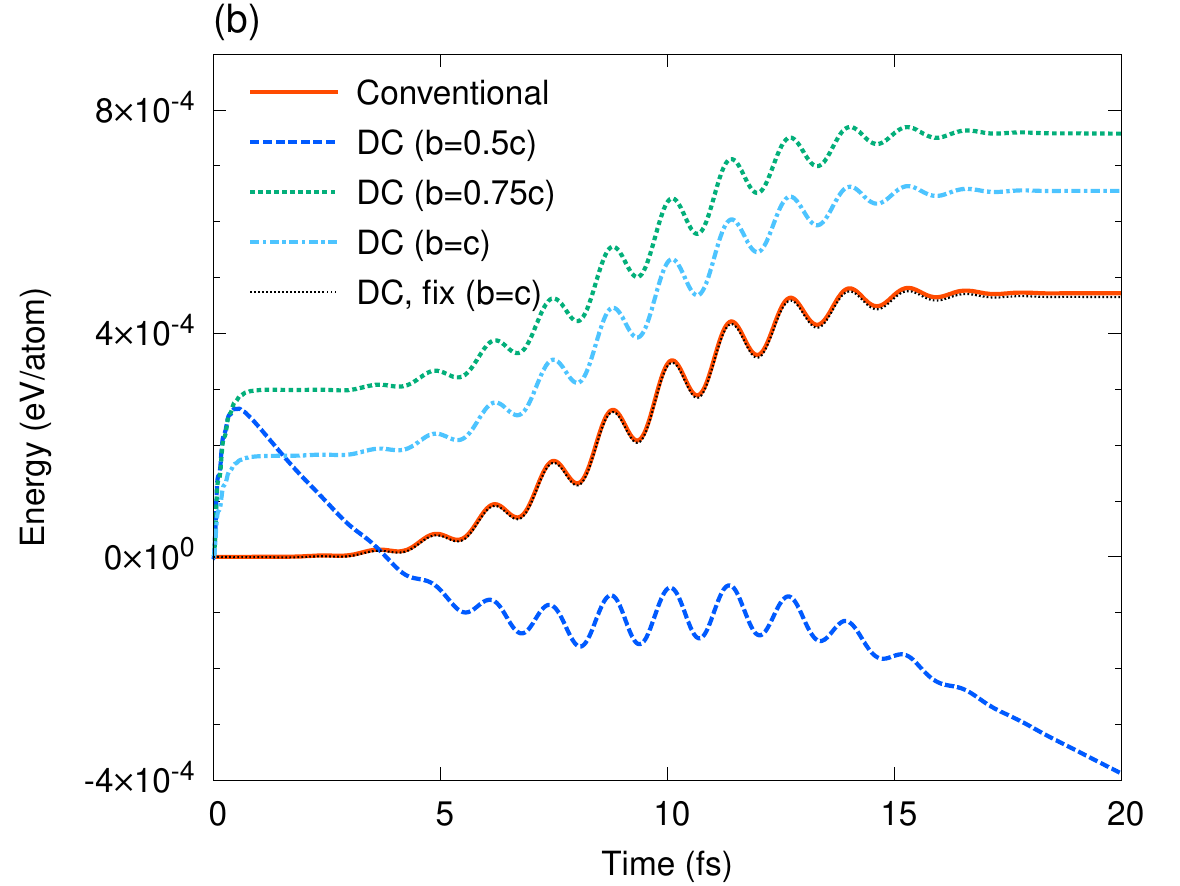}
    \end{tabular}
    \caption{\label{fig:aSi_w}  
     Current density (a) and excitation energy (b) in the 512-atom amorphous Si system from real-time TDDFT calculations for a laser pulse with $I = 10^9$~W/cm$^2$, $\hbar\omega = 1.55$~eV, and $T = 20$~fs, using different initial-state preparation methods. 
     The red solid line corresponds to the initial state obtained from conventional DFT. 
     The other curves correspond to initial states obtained from DC-DFT with buffer thickness $b$ ($c$ is the side length of the core domain).
     The thin black dotted line denotes the fixed-density DC calculation for $b = c$.
    }
\end{figure}

We assess the accuracy of real-time TDDFT calculations initialized with global KS orbitals prepared by DC-DFT using an amorphous Si system.
From the nearsightedness principle of DFT~\cite{Kohn1996,Prodan2006}, DC-DFT can perform well for gapped and disordered systems, where correlations in the density matrix decay exponentially. 
The deviation of DC-DFT from conventional DFT arises from truncating the electron density at short distances. The resulting accuracy can be improved by increasing the buffer thickness $b$. 
However, increasing $b$ also leads to a higher computational cost, resulting in a trade-off between accuracy and efficiency.
Herein, we focus on amorphous Si, an important disordered system for practical applications. We systematically evaluate how the accuracy of real-time TDDFT calculations initialized with DC-DFT depends on the buffer thickness $b$.

We consider a 512-atom amorphous Si system whose atomic coordinates are taken from Ref.~\cite{Deringer2018}.  
The simulation cell is a cube with side length $L$ = 22~{\AA}, and the real-space grid spacing is set to $L/80 = 0.27$~{\AA}. 
In the DC-DFT calculations, the simulation cell is divided into $4 \times 4 \times 4$ fragments so that the side length of each core region is set to $c = L/4$.

Figures~\ref{fig:aSi_w}(a) and \ref{fig:aSi_w}(b) show the current density and excitation energy, respectively, for a laser pulse with intensity $I$ = $10^9$~W/cm$^2$, photon energy $\hbar\omega$ = 1.55~eV, and pulse duration $T$ = 20~fs. 
This intensity corresponds to an extremely weak field in the linear-response regime.
The red solid line represents the real-time TDDFT calculation initialized with the KS orbitals obtained from conventional DFT, whereas the other curves correspond to calculations initialized with the KS orbitals obtained from DC-DFT. 
The blue dashed, green dotted, and light-blue dash-dotted lines correspond to buffer thicknesses $b = c/2$, $3c/4$, and $c$, respectively. 
In addition, the thin black dotted line represents a calculation for the DC-DFT case with $b = c$ in which the total electron density $\rho_{\rm tot}(\mathbf{r})$ is fixed to that obtained from conventional DFT (fixed-density DC). 
Because the total density is fixed to the reference value in this calculation, we can isolate and assess only the accuracy of the DC-LCFO method used to construct the global KS orbitals.

The current density shown in Fig.~\ref{fig:aSi_w}(a) is in overall agreement for all cases, although some noise is observed around $t = 0$ in the DC-DFT results.
In the fixed-density DC case, the result agrees perfectly with the conventional one. 
By contrast, the excitation energy in Fig.~\ref{fig:aSi_w}(b) shows a much larger deviation of the DC-DFT results from the conventional result.
Because the intensity $I$ = $10^9$~W/cm$^2$ is extremely small, the energy shift originating from the electron density error specific to the DC-DFT SCF loop appears large.
This energy shift arises because the electron density in the initial state of the real-time TDDFT calculation is not fully converged in the SCF procedure, owing to the error of DC-DFT.
We confirm that increasing the buffer thickness $b$ systematically improves the accuracy of the electron density and correspondingly reduces the error in the excitation energy.
Furthermore, the fixed-density DC result agrees with the conventional one, confirming that the actual source of the error is the convergence of the electron density. 
These results demonstrate that the accuracy of the DC-LCFO method is sufficient for practical use. They further show that the remaining error can be systematically reduced by improving the accuracy of the SCF part of the DC-DFT calculation.

\begin{figure}
    \centering
    \begin{tabular}{c}
    \includegraphics[keepaspectratio,width=8cm]{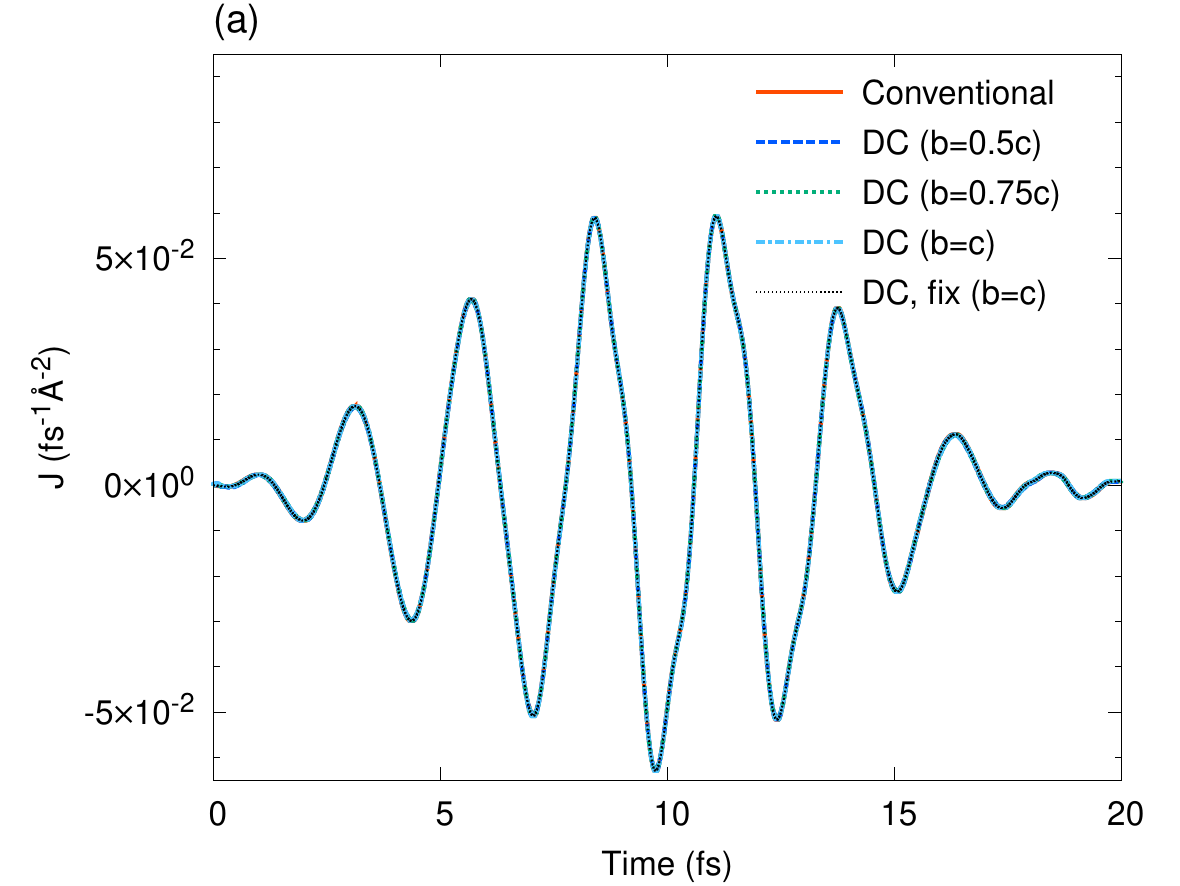}\\
    \includegraphics[keepaspectratio,width=8cm]{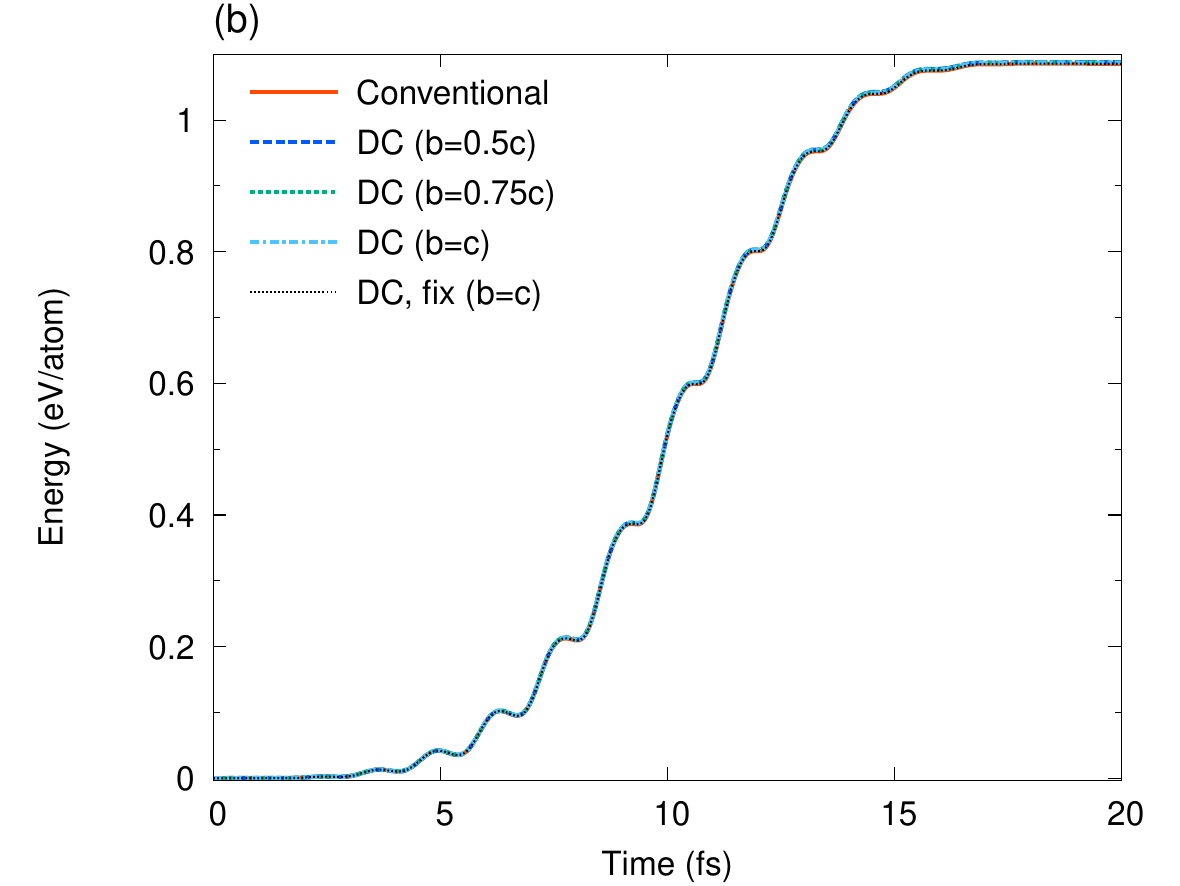}
    \end{tabular}
    \caption{\label{fig:aSi_s}  
    Same as Fig.~\ref{fig:aSi_w}, but for an intensity of $I = 10^{12}$~W/cm$^2$.
    }
\end{figure}

Figure~\ref{fig:aSi_s} shows the same quantities as in Fig.~\ref{fig:aSi_w}, but for a laser intensity of $I = 10^{12}$~W/cm$^2$. 
This corresponds to calculations under an intense laser pulse that induces nonlinear excitation phenomena, such as high-harmonic generation. 
In this case, the error originating from DC-DFT becomes small and the current and excitation energy agree almost perfectly for all cases considered.
This indicates that for real-time dynamical simulations driven by intense laser pulses, the initial states prepared by DC-DFT have sufficient accuracy.

\begin{figure}
    \centering
    \begin{tabular}{c}
    \includegraphics[keepaspectratio,width=8cm]{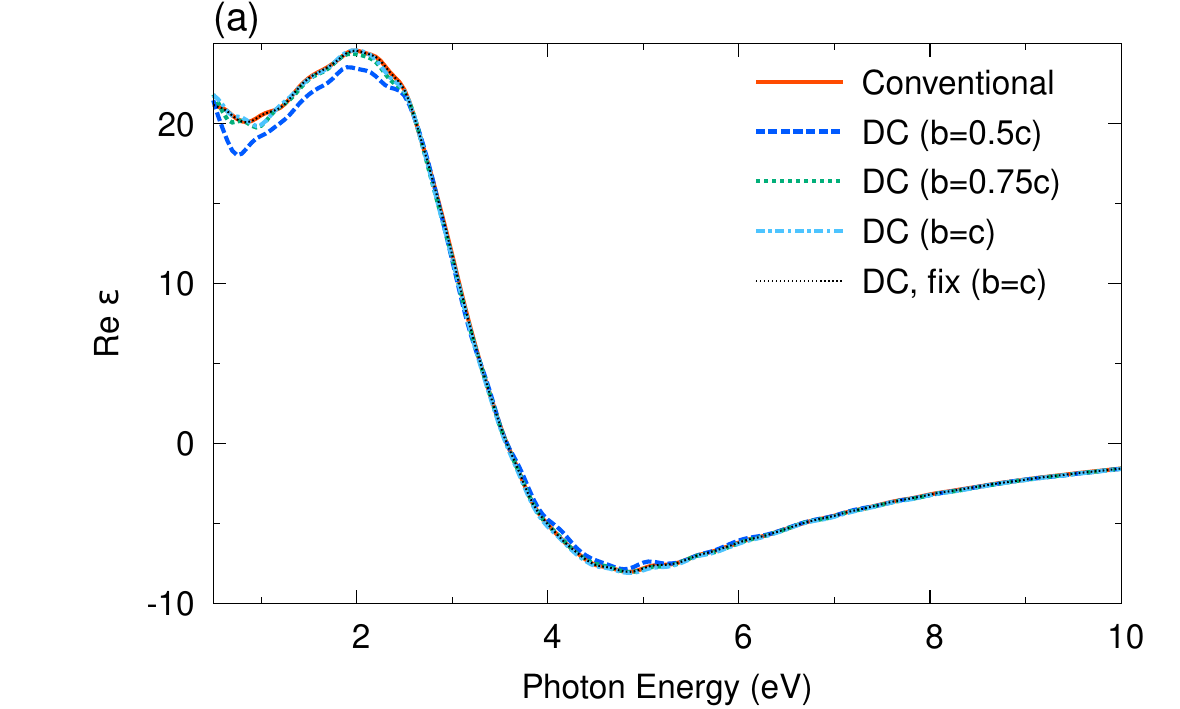}\\
    \includegraphics[keepaspectratio,width=8cm]{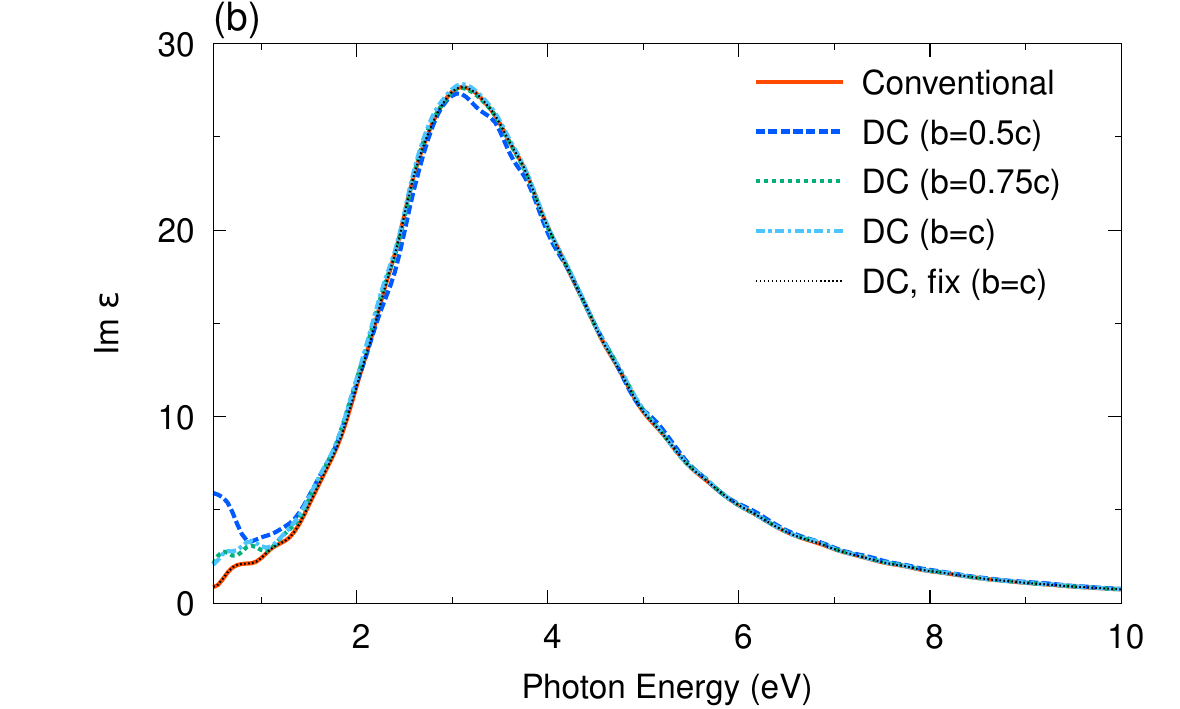}
    \end{tabular}
    \caption{\label{fig:aSi_lr}  
    Real part (a) and imaginary part (b) of the dielectric function of the 512-atom amorphous Si system.
    }
\end{figure}

Next, we assess the accuracy of frequency-domain analysis. 
Figure~\ref{fig:aSi_lr} shows the real part (a) and the imaginary part (b) of the dielectric function of amorphous Si obtained from real-time linear-response calculations.
In these calculations, an impulsive electric field in the linear-response regime is applied at $t=0$ and the dielectric function is obtained by Fourier-transforming the current calculated up to $t=20$~fs. 
The results obtained with conventional DFT and DC-DFT are in overall agreement, although deviations are observed around zero frequency. 
The error also tends to decrease as the buffer thickness $b$ increases.
In addition, the fixed-density DC result agrees almost perfectly with the conventional result.

\begin{figure}
    \centering
    \includegraphics[keepaspectratio,width=8cm]{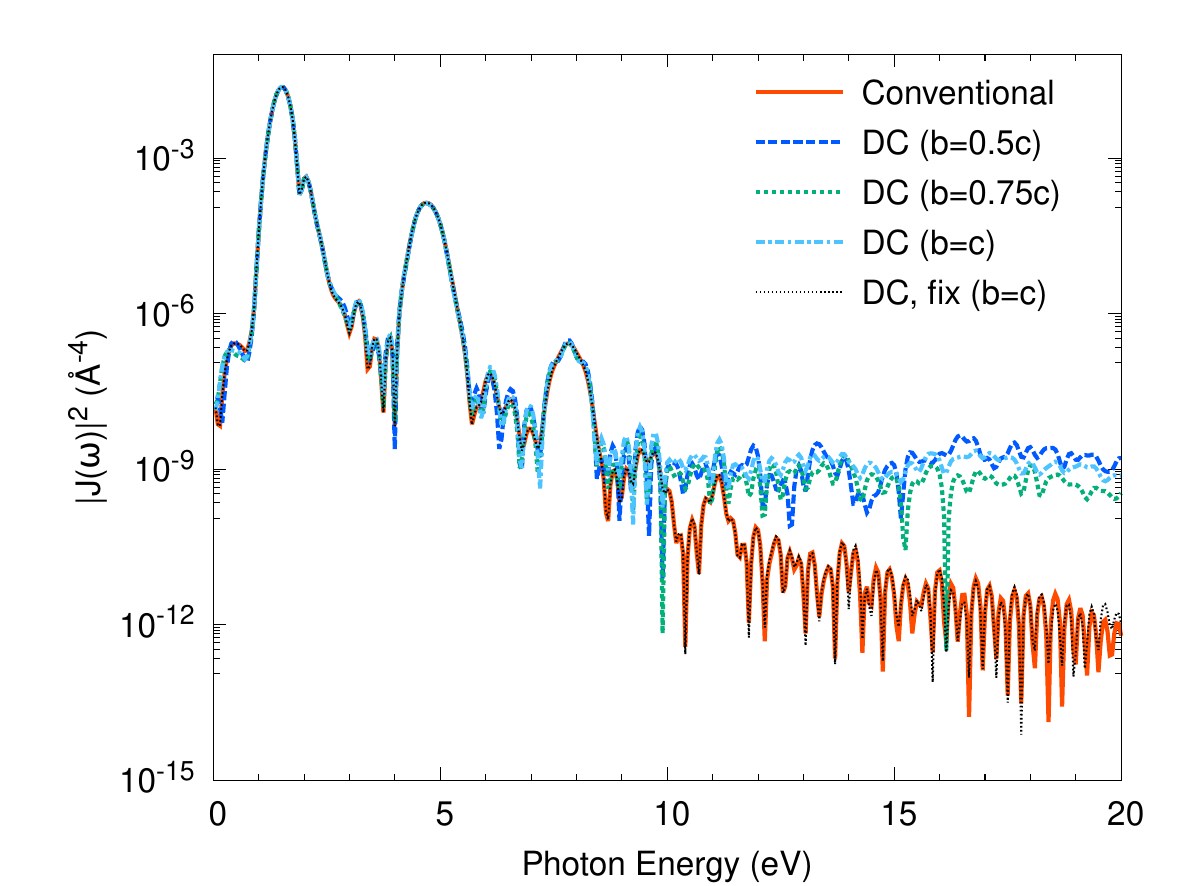}
    \caption{\label{fig:aSi_hhg}  
    High-harmonic spectrum calculated from the current density shown in Fig.~\ref{fig:aSi_s}(a).
    }
\end{figure}

Figure~\ref{fig:aSi_hhg} shows the high-harmonic spectrum obtained by Fourier-transforming the current calculated in Fig.~\ref{fig:aSi_s}(a) and plotting the squared magnitude as a function of photon energy. 
Below 10~eV, the first-, third-, and fifth-order harmonic peaks can be identified from left to right. 
In the DC-DFT cases, notable background noise appears above 10~eV, which buries the seventh harmonic around 11~eV that is visible in the conventional result. 
By contrast, the fixed-density DC result agrees almost perfectly with the conventional result even in the high-frequency region.

These results indicate that DC-DFT initial states are generally accurate enough for practical analyses in the frequency domain as well. 
However, in calculations that are particularly sensitive to background noise in the current, such as the low-frequency region of linear response and the high-frequency region of high-harmonic spectra, the electron density error specific to DC-DFT becomes a limiting factor. 
For calculations requiring such high accuracy, the DC-DFT initial states need to be connected to a conventional DFT calculation to remove the noise. 
In this case, the wavefunctions from conventional DFT do not need to be stored because sufficient accuracy can be obtained by storing only the electron density and reconstructing the global KS orbitals via fixed-density DC calculation. 
This approach is especially useful when storage capacity is a concern because it avoids the need to generate extremely large wavefunction files.

\subsection{Bulk H$_2$O liquid}

\begin{figure}
    \centering
    \begin{tabular}{c}
    \includegraphics[keepaspectratio,width=8cm]{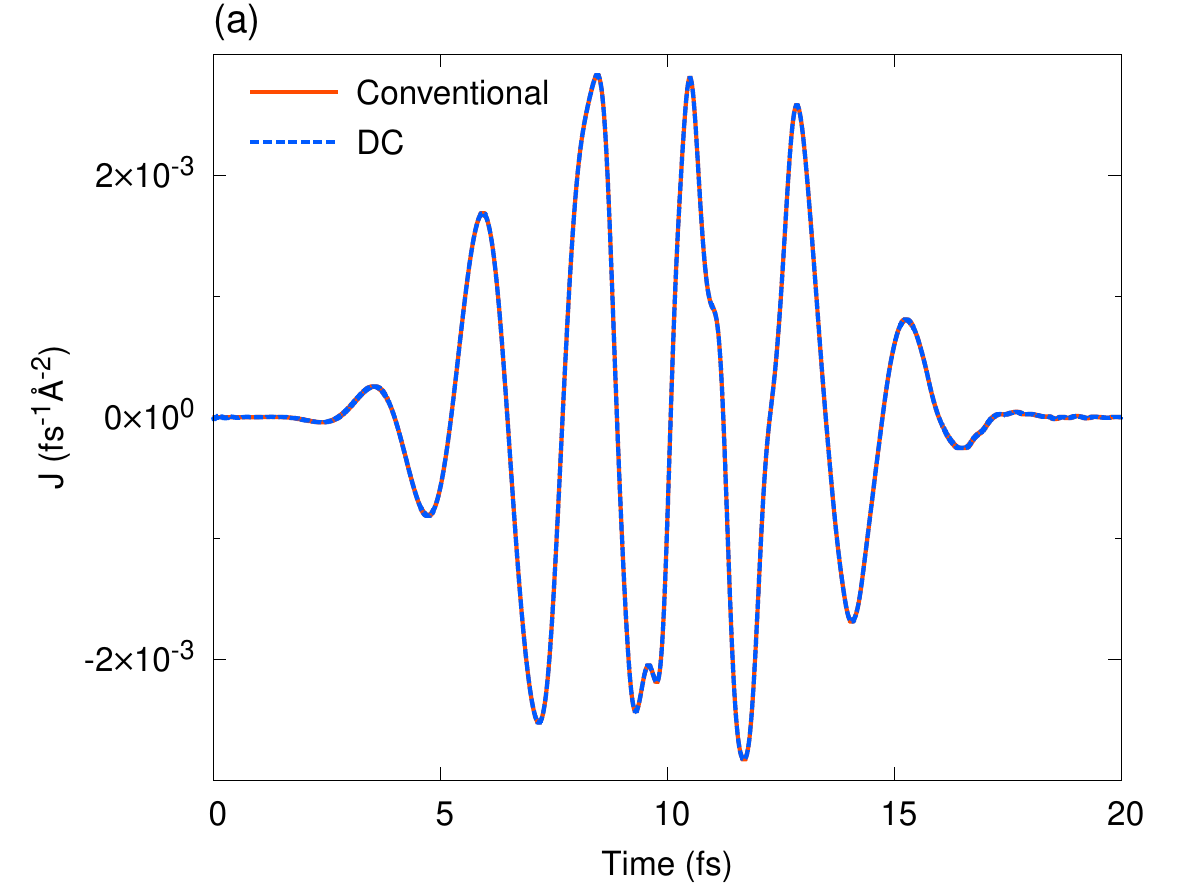}\\
    \includegraphics[keepaspectratio,width=8cm]{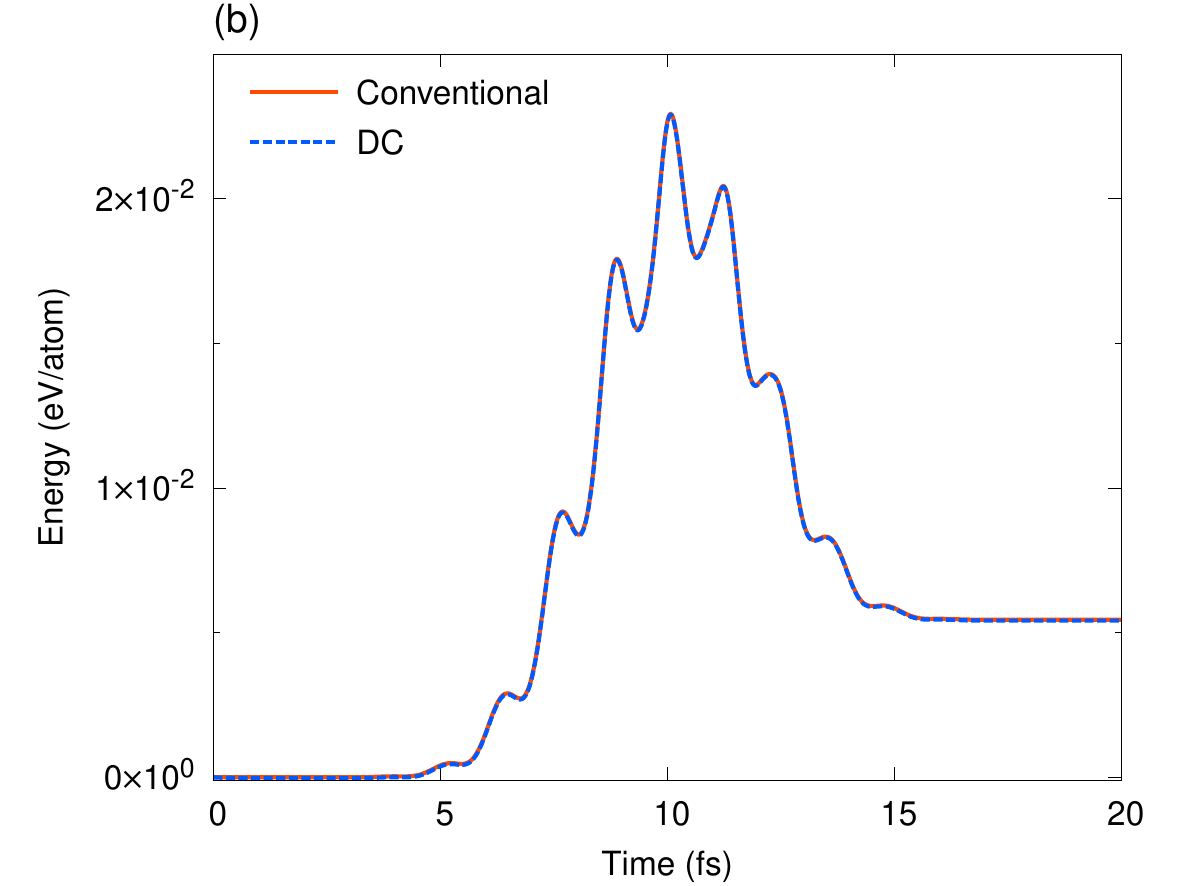}\\
    \includegraphics[keepaspectratio,width=8cm]{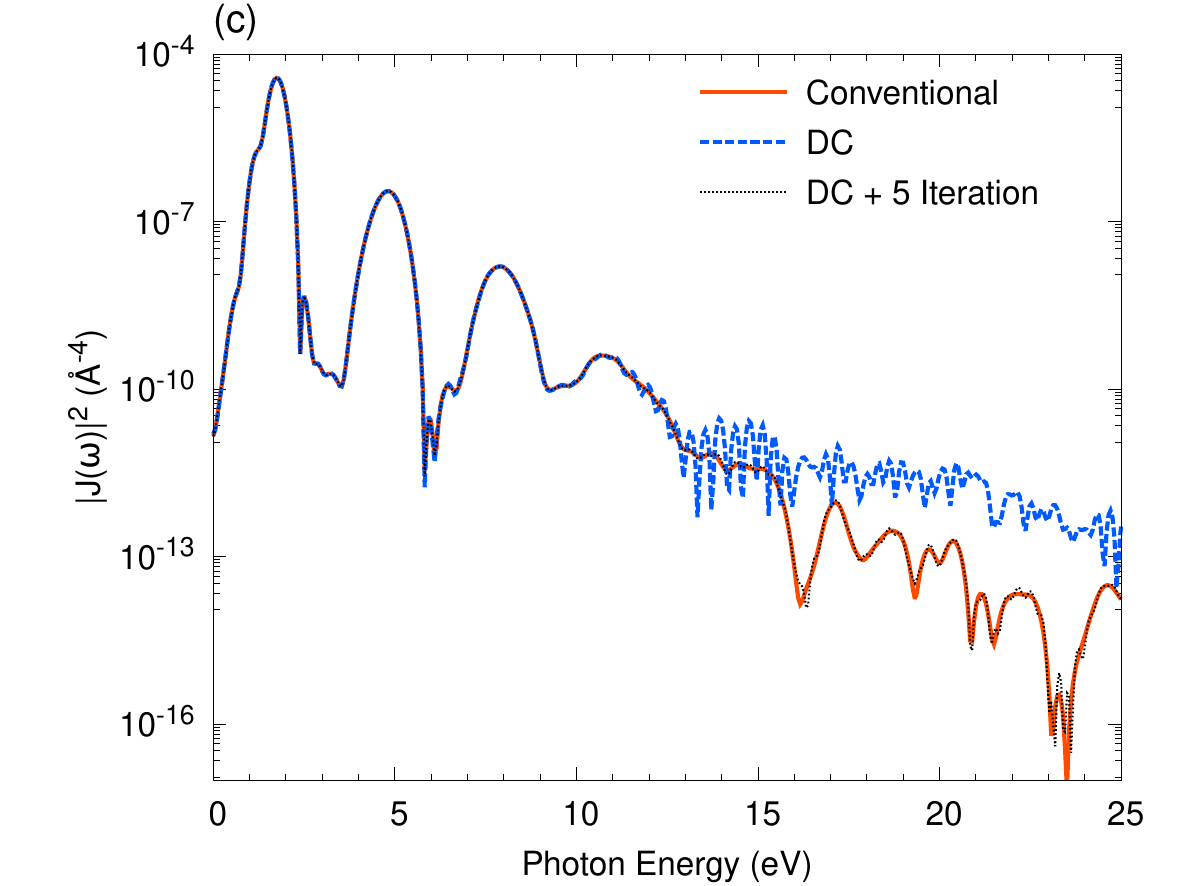}
    \end{tabular}
    \caption{\label{fig:h2o}  
    Current density (a), excitation energy (b), and high-harmonic spectrum (c) for the 4,134-atom liquid-water system from real-time TDDFT calculations with different initial-state preparation methods for a laser pulse with $I = 10^{13}$~W/cm$^2$, $\hbar\omega = 1.55$~eV, and $T = 20$~fs. 
    The red solid and blue dashed lines represent results from conventional DFT and DC-DFT, respectively. 
    The thin black dotted line in panel (c) corresponds to five additional SCF iterations of conventional DFT starting from the DC-DFT initial states.
    }
\end{figure}

Figure~\ref{fig:h2o} shows the current density (a), excitation energy (b), and high-harmonic spectrum obtained from the current density (c) for liquid water.
In this calculation, we use a bulk cubic periodic cell with side length 35~{\AA} containing 4,134 atoms, i.e., 1,378 H$_2$O molecules. 
The atomic structure is obtained using GROMACS code~\cite{Abraham2015,GROMACS}.
The applied laser pulse has an intensity of $I = 10^{13}$~W/cm$^2$, a photon energy of $\hbar\omega = 1.55$~eV, and a pulse duration of $T = 20$~fs. 
The real-space grid spacing is set to $35/160 = 0.22$~{\AA}. 
In the DC-DFT calculation, we use an $8 \times 8 \times 8$ fragment decomposition with a buffer thickness of $b = 3.5$~{\AA}. 
The red solid and blue dashed lines represent the real-time TDDFT results obtained using initial states prepared by conventional DFT and DC-DFT, respectively.

Figure~\ref{fig:h2o}(a) shows that the current exhibits strongly nonlinear behavior under an extremely intense electric field. 
In Figs.~\ref{fig:h2o}(a) and \ref{fig:h2o}(b), the results of conventional DFT and DC-DFT agree almost perfectly. 
In the high-harmonic spectrum shown in Fig.~\ref{fig:h2o}(c), the two results also agree well in the low-frequency region below 10~eV, whereas deviations appear in the higher-frequency region. 
These trends are consistent with those observed for amorphous Si under intense laser irradiation. 
Therefore, we performed an additional conventional DFT calculation starting from the DC-DFT initial states, with only five SCF iterations, and the result of the subsequent real-time TDDFT calculation is shown by the thin black dotted line in Fig.~\ref{fig:h2o}(c). 
The noise originating from the density error in DC-DFT is removed, and the result agrees almost perfectly with that of conventional DFT, even in the high-frequency region. 
We further confirmed that after five more iterations, i.e., 10 SCF iterations, the result becomes indistinguishable from the conventional DFT result.

The wall-clock times for DFT, DC-DFT, and TDDFT were 4.3~h, 3~min, and 5.7~h, respectively, on 512 Fugaku nodes.
In the DFT and DC-DFT calculations, convergence was judged using the criterion $\int d^3r |\rho_i({\mathbf{r}})-\rho_{i-1}({\mathbf{r}})|/N_e < 10^{-9}$, where $i$ denotes the iteration index. 
Convergence was reached in 166 iterations for DFT and 152 iterations for DC-DFT.
Of the total DC-DFT time of 183~s, 67~s was spent in the SCF loop, 53~s in the preparation of the basis functions for the DC-LCFO method, and 62~s in the diagonalization of the Hamiltonian matrix.
The diagonalization was performed using EigenExa, and the matrix dimension was 56,163.
This corresponds to 13.6 basis functions per atom, which is comparable to the number of basis functions required in Ref.~\cite{Yamada2017}.

In this calculation, replacing DFT with DC-DFT reduced the computational time for initial-state preparation from 4.3~h to 3~min. 
Even when 10 additional DFT iterations are performed starting from the DC-DFT initial states, the required computation time remains less than 20~min. 
This practically relevant example demonstrates that the DC-DFT implementation in SALMON can drastically reduce the computational cost of initial-state preparation for large-scale TDDFT simulations.

For the present 4,134-atom system, the cost of conventional DFT and that of TDDFT propagation are comparable.
Therefore, if only a single real-time simulation is required, replacing the conventional DFT step with DC-DFT provides a moderate advantage. 
However, when several TDDFT calculations are performed from the same ground state, for example, with different laser intensities, polarizations, or pulse shapes, the cost of a single conventional DFT calculation can be amortized over multiple propagations. 
In such a situation, especially when the high accuracy of the initial density is required, the practical advantage of DC-DFT becomes less pronounced for a system of this size. 
This balance changes rapidly for larger systems. As the conventional DFT preparation scales approximately as \(O(N^3)\), whereas the real-time TDDFT propagation scales as \(O(N^2)\), doubling the system size makes the DFT stage grow faster than the propagation stage. 
Thus, for systems of ~8,000 atoms, DC-DFT can provide a clear advantage even when several real-time propagations are performed. For systems of ~16,000 atoms or larger, the conventional DFT ground-state calculation would become a dominant cost and the DC-DFT-based workflow should be substantially more favorable.

The sizes of the files \path{basis_functions.bin}, \path{hamiltonian_local.bin}, \path{wavefunctions.bin}, and \path{rgrid_index.bin} in each fragment directory were 16, 13, 12, and 0.6 MB, respectively. 
As the calculation comprises 512 fragments in total, the total size of the \path{data_dcdft} directory amounts to 20 GB. 
By contrast, the wavefunction file stored in a conventional DFT calculation is 170 GB. 
This demonstrates that DC-DFT offers a substantial advantage in terms of file size.


\section{Conclusion}
\label{sec:conclusion}
Herein, we developed a practical route toward large-scale simulations of strongly driven electron dynamics by combining linear-scaling ground-state preparation with the established grid-based real-time TDDFT framework of SALMON.
We have implemented a DC-DFT workflow in SALMON and established its direct connection to conventional real-time TDDFT calculations. This connection is achieved through a postprocessing method that reconstructs global KS orbitals. 
The present implementation combines four steps: DC-DFT-based SCF calculations, construction of orthonormal basis functions from localized KS orbitals used in the DC method, assembly and diagonalization of a reduced Hamiltonian matrix, and reconstruction of global wavefunctions. These reconstructed wavefunctions can then be used as initial states for standard real-time propagation. 
Taken together, this provides a practical, end-to-end workflow for accelerating large-scale electron-dynamics simulations in SALMON.

The performance benchmarks demonstrate that the DC-DFT-based SCF procedure exhibits linear-scaling behavior on the Fugaku supercomputer. They further show that the overall workflow can be applied to large-scale systems. 
Furthermore, using a 512-atom amorphous Si system and a 4,134-atom liquid-water system, we assessed the accuracy of physical observables obtained from real-time TDDFT calculations initialized with DC-DFT. We also identified the regimes in which practically sufficient accuracy is achieved. 
In particular, for the liquid-water system, we showed that the computational time for initial-state preparation can be reduced from 4.3~h to 3~min compared with the conventional approach. The file size required for constructing the KS orbitals can likewise be reduced relative to the conventional approach, from 170 GB to 20 GB. 
These examples illustrate the practical usage and advantages of the DC-DFT workflow implemented in SALMON.

The present implementation is not intended to replace localized-basis real-time TDDFT approaches. Instead, it provides a complementary strategy that retains the existing real-space description of highly excited and spatially extended electronic states while removing the bottleneck of ground-state preparation. This capability should be particularly valuable for simulations of strong-field and nonequilibrium phenomena in realistic systems containing thousands to tens of thousands of atoms.

\section{Acknowledgements}
The authors thank Dr. Junichi Iwata (Quemix Inc.) for the fruitful discussions.
This research was supported by a grant (No. JPMXS0118067246) from the Japanese Ministry of Education, Culture, Sports, Science and Technology (MEXT) Quantum Leap Flagship Program (Q-LEAP) and by Kakenhi grants‐in‐aid (Nos. 24K01224 and 24K17629) from the Japan Society for the Promotion of Science (JSPS). 
Calculations were performed on the Fugaku supercomputer with support from the HPCI System Research Project (Project IDs: hp240124 and hp250102), the Miyabi supercomputer at the University of Tokyo under a Multidisciplinary Cooperative Research Program of the Center for Computational Sciences at the University of Tsukuba.





\bibliographystyle{elsarticle-num}








\end{document}